\documentclass{sig-alternate}

\usepackage{algorithm}
\usepackage{algorithmic}
\usepackage{amsfonts}
\usepackage{amsmath}
\usepackage{amssymb}
\usepackage{graphics}
\usepackage{graphicx}
\usepackage{hyperref}
\usepackage{latexsym}
\usepackage{mathrsfs}
\usepackage{multirow}
\usepackage{subfigure}
\usepackage{times}
\usepackage{url}

\begin{document}


\title{ChemXSeer Digital Library Gaussian Search}

\numberofauthors{6}

\author{
\alignauthor Shibamouli Lahiri\\
       \affaddr{Computer Science and Engineering}\\
       \affaddr{The Pennsylvania State University}\\
       \affaddr{University Park, PA 16802}\\
       \email{shibamouli@psu.edu}
\alignauthor Juan Pablo Fern\'{a}ndez Ram\'{i}rez\\
       \affaddr{Information Sciences and Technology}\\
       \affaddr{The Pennsylvania State University}\\
       \affaddr{University Park, PA 16802}\\
       \email{juanp\_fernandez@hotmail.com}
\alignauthor Shikha Nangia\\
       \affaddr{Biomedical and Chemical Engineering}\\
       \affaddr{Syracuse University}\\
       \affaddr{Syracuse, NY 13244}\\
       \email{snangia@syr.edu}
\and  
\alignauthor Prasenjit Mitra\\
       \affaddr{Information Sciences and Technology}\\
       \affaddr{The Pennsylvania State University}\\
       \affaddr{University Park, PA 16802}\\
       \email{pmitra@ist.psu.edu}
\alignauthor C. Lee Giles\\
       \affaddr{Information Sciences and Technology}\\
       \affaddr{The Pennsylvania State University}\\
       \affaddr{University Park, PA 16802}\\
       \email{giles@ist.psu.edu}
\alignauthor Karl T. Mueller\\
       \affaddr{Chemistry}\\
       \affaddr{The Pennsylvania State University}\\
       \affaddr{University Park, PA 16802}\\
       \email{ktm2@psu.edu}
}

\maketitle

\begin{abstract}
We report on the Gaussian file search system designed as part of the ChemXSeer digital library. Gaussian files are produced by the Gaussian software~\cite{gaussian}, a software package used for calculating molecular electronic structure and properties. The output files are semi-structured, allowing relatively easy access to the Gaussian attributes and metadata. Our system is currently capable of searching Gaussian documents using a boolean combination of atoms (chemical elements) and attributes. We have also implemented a faceted browsing feature on three important Gaussian attribute types - Basis Set, Job Type and Method Used. The faceted browsing feature enables a user to view and process a smaller, filtered subset of documents.
\end{abstract}




\section{Introduction}
\label{sec:intro}
ChemXSeer is a digital library and data repository for the Chemoinformatics and Computational Chemistry domains~\cite{Mitra:2007:CDL:1317353.1317356}. It currently offers search functionalities on papers and formulae, CHARMM calculation data and Gaussian computation data, and also features a comprehensive search facility on chemical databases. A table search functionality~\cite{Liu:2007:TAT:1255175.1255193}, similar in spirit to the one featured in CiteSeerX\footnote{http://citeseerx.ist.psu.edu/}, is currently under development. Gaussian document search has been a key component of ChemXSeer from its inception. The alpha version of Gaussian search featured a simple query box and an SQL back-end. Here we describe the next generation of Gaussian search\footnote{http://cxs05.ist.psu.edu:8080/ChemXSeerGaussianSearch} which includes a customized user interface for Computational Chemistry researchers, boolean query functionality on a pre-specified set of attributes, and a faceted browsing option over three key attribute types. The current version of Gaussian search is powered by Apache Solr\footnote{http://lucene.apache.org/solr/}, a state-of-the-art open-source enterprise search engine indexer.

The organization of this paper is as follows. In Section~\ref{sec:file_structure}, we give a brief overview of the Gaussian software and Gaussian files, emphasizing the need for a customized search interface rather than a simple one. Description of the search interface appears in Section~\ref{sec:system_description}, followed by a brief sketch of related work in Section~\ref{sec:related_work}. We conclude in Section~\ref{sec:conclusion}, outlining our contributions and providing directions for future improvement.

\section{Gaussian Files}
\label{sec:file_structure}

\begin{figure}
\begin{center}
\includegraphics[width=0.8\linewidth]{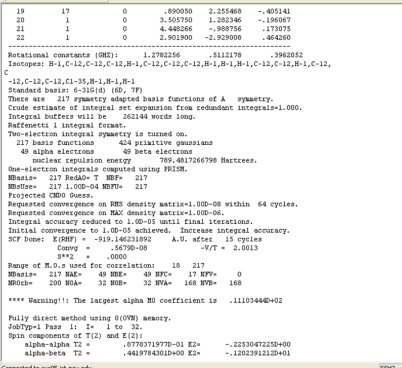}
\end{center}
\caption{Screenshot of a Gaussian document.}
\label{fig:gaussianfile}
\end{figure}

\begin{figure}
\begin{center}
\includegraphics[width=0.6\linewidth]{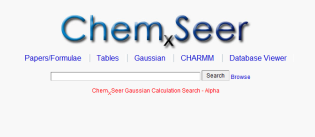}
\end{center}
\caption{First-generation Gaussian query interface.}
\label{fig:oldcxs}
\end{figure}

Computational chemists perform Gaussian calculations to determine properties of a chemical system using a wide array of computational methods. The methods include molecular mechanics, ground state semi-empirical, self-consistent field, and density functional calculations. Computational methods such as these are key to the upsurge of interest in chemical calculations, partly because they allow fast, reliable, and reasonably easy analysis, modeling, and prediction of known and proposed systems (e.g., atoms, molecules, solids, proposed drugs, etc.) under a wide range of physical constraints, and partly because of the availability of well-tested, comprehensive software packages like Gaussian that implement many of these methods with good tradeoff between accuracy and processing time.

The Gaussian software is actually a suite of several different chemical computation models, including packages for molecular mechanics, Hartree-Fock methods, and semi-empirical calculations. While the exact details of the functionalities of this software are beyond the scope of this paper\footnote{For details, please see\\ http://www.gaussian.com/g\_tech/g\_ur/g09help.htm}, we would like the reader to note that each run of the Gaussian software is equivalent to conducting a chemical experiment with certain inputs and under certain physico-chemical conditions. The output of the software consists of a large amount of information returned to the user via the computer console and usually redirected to a suitably-named output file. We are interested in these output files, henceforth referred to as ``Gaussian files'' or ``Gaussian documents''.

The Gaussian files contain detailed information about the calculations being performed on the system of interest. Although the details of the calculations are essential for the analysis of the system being studied, the output file can be cumbersome to a new user. Each Gaussian file begins with the issued command that initiated a particular calculation, followed by copyright information, memory and hard disk specification, basis set, job type, method used, and several different matrices (e.g., Z-matrix, distance matrix, orientation matrix, etc.). It may also contain other information like rotational constants, trust radius, maximum number of steps, and steps in a particular run. Gaussian files are semi-structured (Figure~\ref{fig:gaussianfile}) in the sense that these parameters tend to appear in a particular order or with explicit markups.

Since Gaussian files are important to the design, testing and prediction of new chemical systems, ChemXSeer had integrated a search functionality on these files. The alpha version of Gaussian search interface only consisted of a simple query box (Figure~\ref{fig:oldcxs}), and the back-end of the search engine was an SQL database that stored data extracted from the Gaussian files. Although simple, the interface allowed users to type in fielded queries and view results in an easy-to-understand format. In the current version, we have retained many aspects of the alpha version, including parts of the search results page and visual representation of individual Gaussian files.

However, our domain experts argued that a more complex interface including faceted search was justified, partly because it eases the task of a researcher by limiting the number of search results to examine, and partly because such interfaces have already been successfully implemented~\cite{doi:10.1021/ci600510j}. A computational chemist usually knows what kinds of parameters he/she is looking for in a Gaussian files database, and therefore it makes sense to refine search results using this information. We identified three important parameters towards this end - Job Type, Method Used and Basis Set. There are other parameters and metadata that we can extract from the Gaussian files, but they are not as important from a domain expert's point of view. These are Charge, Degree of Freedom, Distance Matrix, Energy, Input Orientation, Mulliken Atomic Charge, Multiplicity, Optimized Parameters, Frequencies, Thermo-chemistry, Thermal Energy, Shielding Tensors, Reaction Path, PCM, and Variational Results. Metadata like ID, Title and File Path are used in organizing the search results.

\section{System Description}
\label{sec:system_description}

\begin{figure}
\begin{center}
\includegraphics[width=0.8\linewidth]{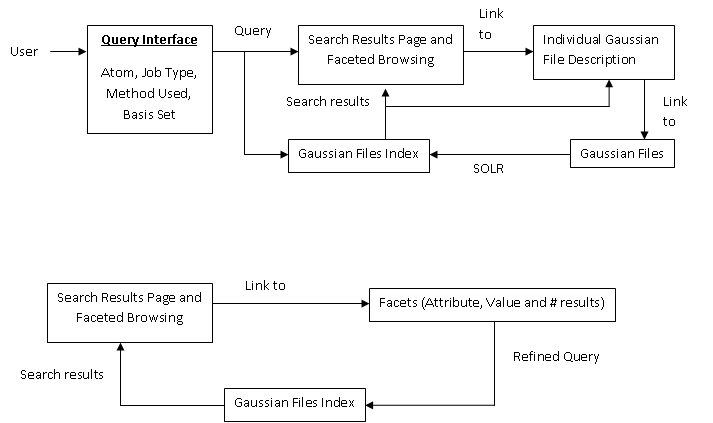}
\end{center}
\caption{Gaussian search system architecture.}
\label{fig:blockdiagram}
\end{figure}

The basic query to the Gaussian search system is an atom (i.e., element) or a collection of atoms. The system returns all Gaussian files containing those atoms. However, as experienced by researchers, such basic queries often return a large number of search results, many of which are not relevant. While we can think of improving the ranking of search results in tune with traditional information retrieval research, domain experts have informed us that since Gaussian files are semi-structured, a faceted browsing option would be more appropriate. It remains open, however, whether ranking within each facet could be improved. Currently we rank the search results by their external IDs, because our domain experts were not overly concerned with the ranking.

The system architecture is given in Figure~\ref{fig:blockdiagram}. Figure~\ref{fig:blockdiagram} has three principal components - the query interface, the search results page and the Gaussian file description page. The user supplies a query using the query interface, consisting of atoms (mandatory field), method used, job type and basis set. The last three fields are optional, and can be combined in boolean AND/OR fashion. The boolean query goes to the Gaussian document index, which in turn returns on the search results page all Gaussian files satisfying the boolean query. The search results page contains links to individual Gaussian file descriptions, which in turn link to the actual Gaussian documents. Figure~\ref{fig:blockdiagram} also indicates that the index was generated from Gaussian documents using Apache Solr.

The lower section of Figure~\ref{fig:blockdiagram} explains the faceted browsing part. Facets are created based on three attributes - job type, method used, and basis set. Each facet link consists of an attribute, its value, and the number of search results under the current set that satisfy this value. The search results page contains links to different values of the attributes. When the user clicks on such a link, a refined query is sent to the Gaussian document index and the resulting smaller set of search results is returned.

\begin{figure}
\begin{center}
\includegraphics[width=0.8\linewidth]{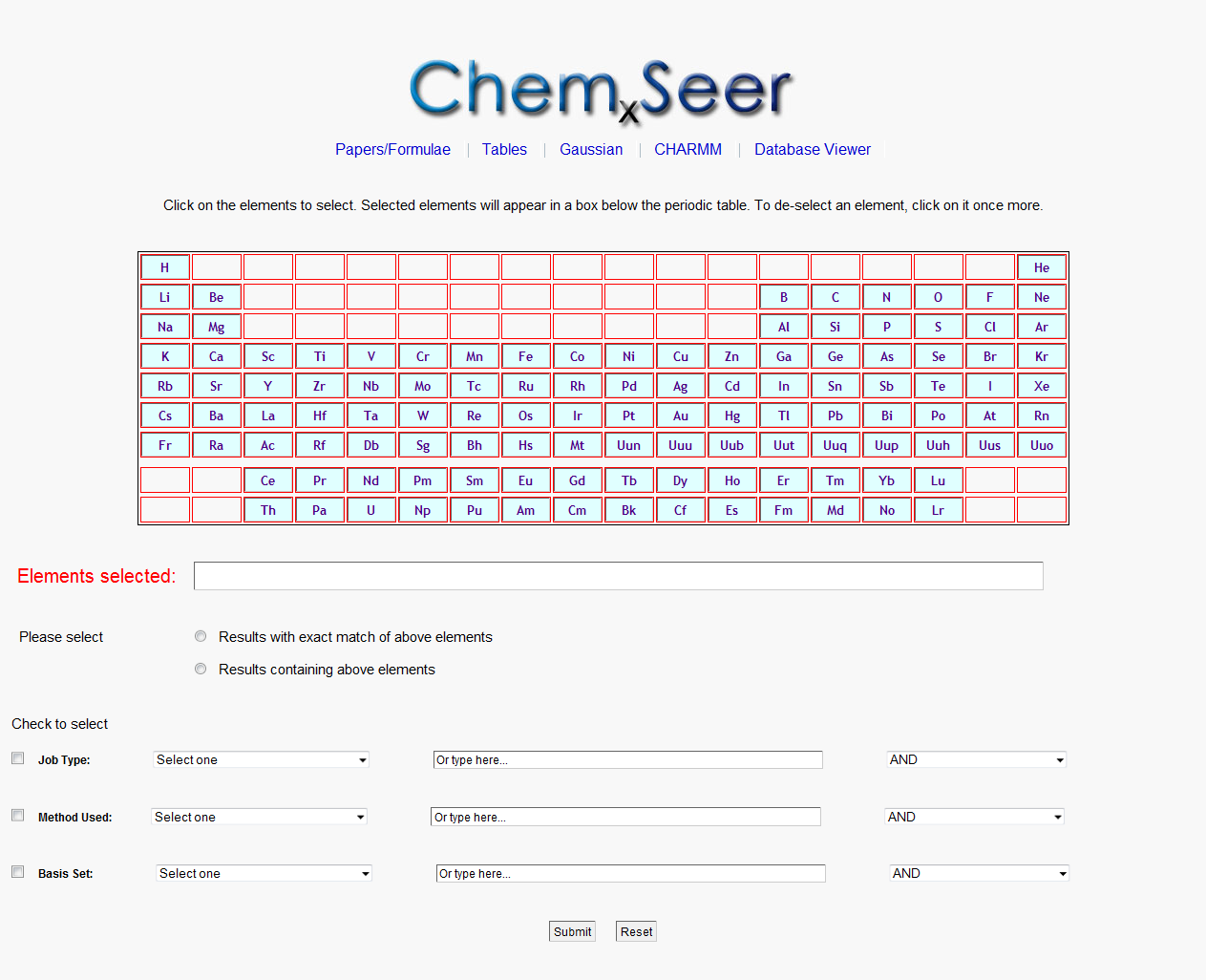}
\end{center}
\caption{Gaussian query interface.}
\label{fig:newcxs}
\end{figure}

\begin{table}
\caption{Gaussian Attribute Categories}
\begin{center}
\begin{tabular}{|c|c|c|}
\hline
\textbf{Job Type} & \textbf{Method Used} & \textbf{Basis Set}\\
\hline
Any & Any & Any \\
Single Point & Semi-empirical & gen\\
Opt & Molecular Mechanics &\\
Freq & Hartree-Fock &\\
IRC & MP Methods &\\
IRCMax & DFT Methods &\\
Force & Multilevel Methods &\\
ONIOM & CI Methods &\\
ADMP & Coupled Cluster Methods &\\
BOMD & CASSCF &\\
Scan & BD &\\
PBC & OVGF &\\
SCRF & Huckel &\\
NMR & Extended Huckel &\\
& GVB &\\
& CBS Methods &\\
\hline
\end{tabular}
\end{center}
\label{tab:g_att_cat}
\end{table}

The implementation of our query interface (Figure~\ref{fig:newcxs}) was inspired partly by the EMSL Basis Set Exchange interface\footnote{https://bse.pnl.gov/bse/portal}, and partly by the requirements mentioned by our domain experts. Our interface features a periodic table of elements, where users can click to select and de-select each element (atom) individually. The selected elements appear together in the textbox at the bottom of the table. Users can specify whether they want search results that contain only the selected elements - no more and no less, or whether they want search results that contain the selected elements as well as other elements. After selecting elements, users can optionally select Job Type, Method Used, and Basis Set from the drop-down menus provided. They can also directly type in the desired values for these attributes in the textboxes. Finally, they can specify AND/OR from another drop-down menu. The default option is AND. Fourteen Job Type categories (values), sixteen Method Used categories, and two Basis Set categories are provided in the drop-down menus. These categories are given in Table~\ref{tab:g_att_cat}. Each category has several sub-categories that are dealt with by the search system. For example, if a user specifies ``Hartree-Fock'' as the Method Used category, the system will search for four sub-categories of Hartree-Fock - hf, rhf, rohf and uhf. These sub-categories were specified by our domain experts. A sample of Method Used sub-categories is given in Table~\ref{tab:sub_cat}. Table~\ref{tab:sub_cat} shows the sub-categories for three Method Used categories - Molecular Mechanics, CI Methods, and CBS Methods. For the Basis Set attribute there are many categories, but only two options are provided in the drop-down menu to keep it short and simple. Users can type in the category (e.g., 3-21G*) in the textbox provided.

\begin{figure}
\begin{center}
\includegraphics[width=0.8\linewidth]{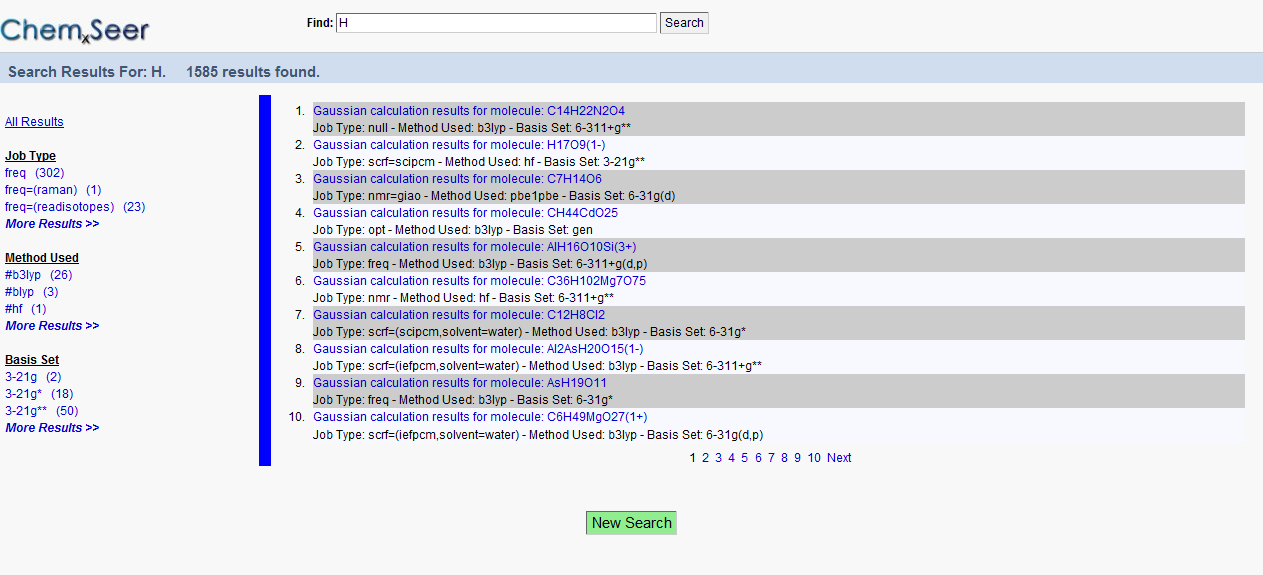}
\end{center}
\caption{A search results page.}
\label{fig:searchresultspage}
\end{figure}

\begin{figure}
\begin{center}
\includegraphics[width=0.8\linewidth]{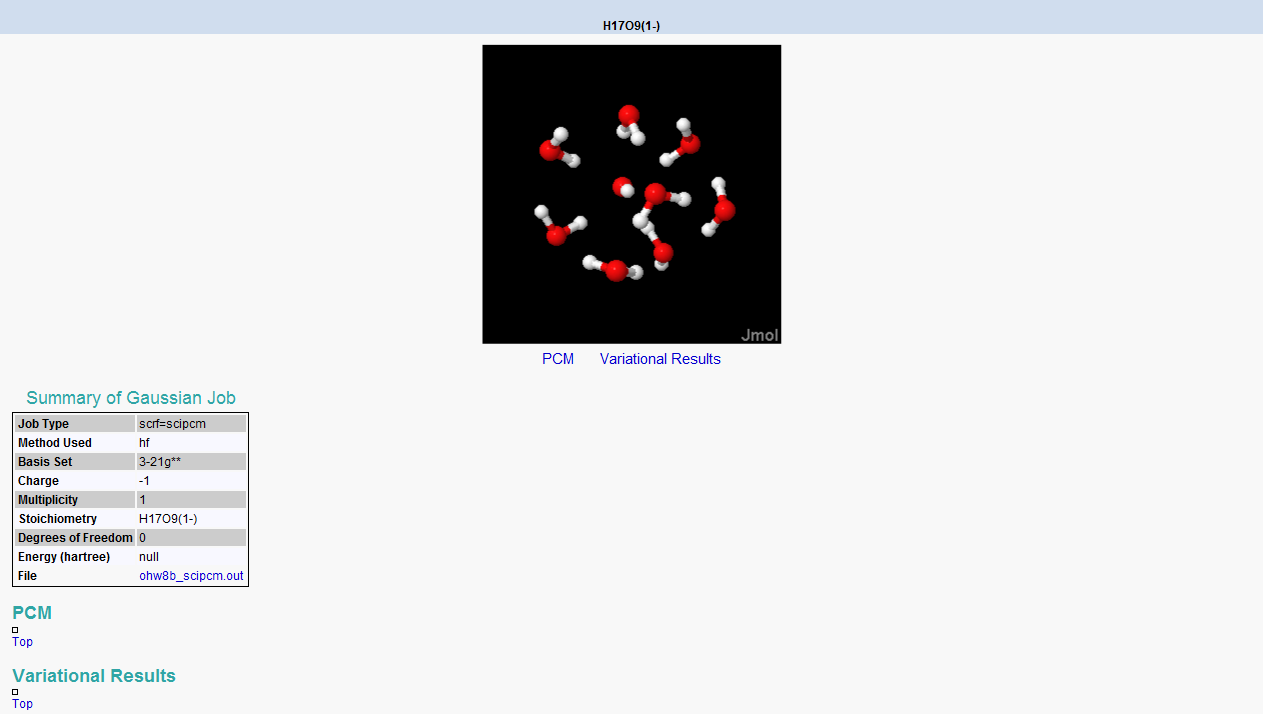}
\end{center}
\caption{A Gaussian file description.}
\label{fig:gaussiandetails}
\end{figure}

Ten search results are shown in one search results page (Figure~\ref{fig:searchresultspage}) with the total number of results shown at the top. Note that the left part of the search results page (Figure~\ref{fig:searchresultspage}) contains links for faceted browsing, and the right part contains the actual results. Each search result consists of a link to the corresponding Gaussian file description and a one-line summary of the file containing attribute information. The Gaussian file description (Figure~\ref{fig:gaussiandetails}) consists of a Jmol~\cite{Hanson:kk5066} rendering of the system being studied, followed by a summary of the Gaussian job and information about attributes extracted from the file. The summary contains a link to the Gaussian document (Figure~\ref{fig:gaussiandetails}). Currently we have indexed 2148 documents.

The faceted browsing section (left half of Figure~\ref{fig:searchresultspage}) follows the architectural specification of Figure~\ref{fig:blockdiagram}. Users can refine search results any time simply by clicking on a particular attribute category. An ``All Results'' link has been provided to help users quickly find the original set of results. Anecdotal evidence from our domain experts suggests that the faceted browsing feature has been able to significantly cut down on the number of search results to examine, thereby saving a considerable amount of time on the part of a Computational Chemistry researcher. Moreover, since each facet link gives the number of search results to examine for a particular attribute category, a user can readily obtain a visual appreciation of the distribution of search results across different attribute categories for a single query.

The core search and indexing functionality of Gaussian search is currently provided by Apache Solr, an open-source state-of-the-art enterprise search server designed to handle, among other things, faceted search, boolean queries, and multivalued attributes. In our case, atoms (chemical elements) in a Gaussian document comprise a multivalued attribute. Each Gaussian document was converted by our home-grown metadata extractor into an XML-style file suitable for ingestion to Solr. The selection of Solr as the back-end platform for this system was partly motivated by the need to integrate ChemXSeer architecture with SeerSuite\footnote{http://sourceforge.net/projects/citeseerx/}, a package of open-source software tools that powers the CiteSeerX digital library.

\begin{table}
\caption{A sample of Method Used sub-categories}
\begin{center}
\begin{tabular}{|c|c|c|}
\hline
\textbf{Molecular Mechanics} & \textbf{CI Methods} & \textbf{CBS Methods}\\
\hline
amber & cis & cbs-4m \\
drieding & cis(d) & cbs-lq\\
uff & cid & cbs-q\\
& cisd & cbs-qb3\\
& qcisd & cbs-apno\\
& qcisd(t) &\\
& sac-ci &\\
\hline
\end{tabular}
\end{center}
\label{tab:sub_cat}
\end{table}

\section{Related Work}
\label{sec:related_work}
In this section we give a brief sketch of the related work. The importance of using large databases to support chemistry calculations has been illustrated by Feller in~\cite{DBLP:journals/jcc/Feller96}. Schuchardt, et al., describe such a database, the Basis Set Exchange~\cite{doi:10.1021/ci600510j}. Basis Set Exchange helps users find particular basis sets that work on certain collections of atoms, while ChemXSeer lets users search Gaussian files with basis sets as boolean query components.

Among other purely chemistry-domain digital libraries, OREChem ChemXSeer by Li, et al.~\cite{Li:2010:OCS:1816123.1816160} integrates semantic web technology with the basic ChemXSeer framework. The Chemical Education Digital Library~\cite{chemeddl} and the JCE (Journal of Chemical Education) Digital Library~\cite{jcedl} focus on organizing instructional and educational materials in Chemistry. Both these projects are supported by NSF under the National Science Digital Library (NSDL). In contrast with these studies, our focus here is to design a search functionality on Gaussian files that helps domain experts locate attribute information more easily.

\section{Conclusion}
\label{sec:conclusion}
In this paper our contributions are two-fold:
\begin{itemize}
\item design of a new search engine for Computational Chemistry research on documents produced by the widely used Gaussian software, and
\item design of a metadata extractor that sieves out several important attributes from the Gaussian documents, and exports them into Solr-ingestible XML format.
\end{itemize}
Future work consists of integration of documents from the ChemXSeer Digital Library with Gaussian files so that users can have an integrated view of calculations, results, and analysis. The metadata extractor could also be improved. There are a few cases where our metadata extractor could not locate certain attribute values, mainly due to the anomalous placement of those attributes in the Gaussian output files. The structure of these documents appeared inconsistent in certain places. Information extraction techniques may be useful for handling these cases. Another area of potential research is improving the ranking of search results. Although our domain experts were not concerned with ranking, it remains to be seen if combining attribute information can help pull up more relevant files earlier in the ranking. Finally, Section~\ref{sec:file_structure} indicates the presence of several other attributes in the Gaussian documents. It would be interesting to explore whether these attributes are useful and can be leveraged to produce additional relevant information.

\section{Acknowledgments}
\label{sec:ack}

This work was partially supported by the National Science Foundation award CHE-0535656.

\bibliographystyle{abbrv}
\small
\bibliography{jcdl11}  

\end{document}